\documentclass{aa}  
\usepackage{graphicx}
\usepackage{txfonts}
\begin{document}
   \title{A new reduced network to simulate detonations in superbursts from mixed H/He accretors}

   \author{C. No\"el\inst{1}
          \and
          S. Goriely \inst{1}
   	  \and
   	  Y. Busegnies \inst{1}
	  \and M.V. Papalexandris\inst{2}
         }

   \offprints{C. No\"el}

   \institute{Institut d'Astronomie et d'Astrophysique, Universit\'e Libre de Bruxelles,
              Campus plaine CP 226, Boulevard du Triomphe, 1050 Bruxelles, Belgium\\
              \email{cnoel@ulb.ac.be}
	       \and
             D\'epartement de M\'ecanique, Universit\'e catholique de Louvain, 1348 Louvain-la-Neuve, Belgium\\
            }


 
  \abstract
   {}
   {We construct a new reduced nuclear reaction network able to reproduce the energy production due to
   the photo-disintegration of heavy elements such as Ru, which are believed to occur during superbursts 
   in mixed H/He accreting systems. We use this network to simulate a detonation propagation, inside a mixture
   of C/Ru.
   }
   {As our reference, we use a full nuclear reaction network, including 14758 reactions on 1381 nuclides.
   Until the reduced and full networks converge to a good level of accuracy in the energy production rate,
   we iterate between the hydrodynamical simulation, with a given reduced network, and the readjustment of a 
   new reduced network, on the basis of previously derived hydrodynamical profiles.
  }
   {We obtain the thermodynamic state of the material after the passage of the detonation, and the final 
   products of the combustion. Interestingly, we find that all  
  reaction lengths can be resolved in the same simulation. This will enable C/Ru detonations to be more
  easily studied in future multi-dimensional simulations, than pure carbon ones.
  We underline the dependence of the combustion products on the 
  initial mass fraction of Ru. In some cases, a large fraction of heavy nuclei, such as Mo, remains 
  after the passage of the detonation front. In other cases, the ashes are principally composed of iron group elements. }
   {}

   
   \maketitle
%

\section{Introduction}\label{Introduction}

   Superbursts have been discovered by long-term monitoring of the X-ray sky, using instruments such as RXTE and 
   BeppoSAX. Compared to normal type I X-ray bursts, superbursts are 1000 times more energetics, with
   integrated burst-energies of approximately $10^{42}$ ergs; they last 1000 times longer, from hours to
   half a day; and they reoccur every one to a few years.
   They are rare: only 13 events have been identified from 8 sources
    (for reviews see Kuulkers \cite{Kuulkers2004}, Cumming \cite{Cumming2005}, and references therein). 
   They exhibit similarities with  
   normal type I X-ray bursts, such as a rapid rise in the light curve, a quasi-exponential decay, and a hardening of 
   the 
   spectrum during the rise 
   followed by a softening during the decay. These attributes are well represented by a blackbody model, which has 
   an effective temperature 
   that grows 
   during the rise, and decreases during the decay phase (Kuulkers \cite{Kuulkers2004}). This suggests that 
   superbursts, 
   like normal type I bursts, are thermonuclear in origin (Cornelisse et al. \cite{Cornelisse2000}). The current view is that 
   superbursts are due to thermally unstable ignition of $ ^{12}$C at densities of about $10^8$ - 
   $10^9 \, $g cm$^{-3}$ (Cumming \cite{cumming2001}, Strohmayer \& Brown \cite{strohmayer2002}, Cumming \cite{Cumming2005}).
  
  Most superbursts have been observed in systems that accrete a mixture of H and He (Kuulkers \cite{Kuulkers2004}). 
  In these H/He accretors, the $ ^{12}$C mass fraction that remains after the combustion of H and He, via the 
  rp-process (Wallace \& Woosley \cite{Wallace1981}), is unknown (Schatz et al. \cite{Schatz2001}, \cite{Schatz2003b},
  Woosley et al. \cite{Woosley2004}). 
  Cumming \& Bildsten (\cite{cumming2001}) showed that a $ ^{12}$C mass fraction of $0.05-0.1$ was sufficient for an 
  unstable ignition; however, fits to the observed lightcurves suggest that the actual abundance is much larger
  (Cumming et al. \cite{cumming2006}). 
  
  Schatz et al. (\cite{Schatz2003}) 
  showed that, at the high 
  temperatures reached in superbursts ($T > 10^9$K), the photodisintegration of the heavy rp-process ashes 
  influences the energetics of the detonation.
  More details about this interesting mechanism are provided below.
  
  To calculate energy production, a simplified nuclear reaction network is, in general, used in astrophysical
  hydrodynamics simulations (e.g. No\"el et al. \cite{Noel2007}). In this study, we extend such a network
  to include a contribution by rp-process ashes.
  However, full nuclear reaction networks are intractable in hydrodynamical 
  simulations, where some $10^7$ calls to 
  the nuclear reaction 
  network, are required. This would be too time-consuming, in particular 
  in multi-dimensional simulations that will be considered in a future work. For this reason, we constructed 
  a reduced nuclear reaction network. This network and the detonation profiles are described in Sect.
  \ref{networkextension}. Our results are discussed in Sect. \ref{Conclusions}.

\section{Nuclear reaction network extension}\label{networkextension}

  In a mixed H/He accreting system, the detonation that leads to the superburst, is expected to propagate into the ashes of the 
  rp-process, which occurred in the upper atmosphere.
  For simplicity, as in Cumming \& Bildsten (\cite{cumming2001}), we represent the ashes using a fiducial heavy nucleus, 
  $^{96}$Ru \footnote{Choosing only $^{96}$Ru implies na\"ively that the electron captures in 
  the high-density 
  environment have not occured yet and that only $^{96}$Ru is produced during the combustion phases preceding the
  superburst. Future detailled hydrodynamical simulation in 
  X-ray burst may provide us with better initial conditions for the superburst.}.
  We simulate the propagation of a detonation, 
  in a mixture of $^{12}$C and $^{96}$Ru, with mass fractions $X_{^{96}\mbox{Ru}}=1-X_{^{12}\mbox{C}}$, using 
  the hydrodynamical algorithm described in No\"el et al. (\cite{Noel2007}).
  This algorithm solves the adiabatic Euler's equations for compressible, non-viscous gas-dynamics with source terms.
  It is a finite-volume method in the spirit of the original 
  MUSCL scheme of van Leer (\cite{vanLeer1979}). The algorithm is of second-order in the smooth part of the flow, and
  avoids dimensional splitting. Parallelization, which is required to solve computationally, the problem in hand, 
  is based on the mpi library, as described in Deledicque \& Papalexandris 
  (\cite{deledicque2005}). 
  The equation of state accounts for partially degenerate and partially relativistic electrons and positrons. 
  The ions are treated as a Maxwell-Boltzmann gas, and the radiation, considered to be at local thermodynamic 
  equilibrium with the matter, follows the Planck law.
  Coulomb corrections to the equation of state are not included. In our simulations, these corrections would change 
  the thermodynamic quantities of our ideal plasma by less than $2-3\%$ (Fryxell et al.
  \cite{Fryxell2000}), but will be included in future work.
  
  It remains untractable at the present time to perform the hydrodynamical simulation with an extented network,
   so that a reduced network needs to be constructed.
  For this purpose, we first extend the reduced network 
  of thirteen species from $^{12}$C to $^{56}$Ni used in No\"el et al. (\cite{Noel2007}), with the $\alpha$-chain 
  between $ ^{64}$Ni and $^{96}$Ru.
  We adopt the nuclear data   
  from the Brussels nuclear reaction rate library (BRUSLIB), based on published compilations of experimental 
  reaction rates, and on the determination of reaction rates 
  using the statistical Hauser-Feshbach model (Aikawa et al. \cite{Aikawa2005}, Arnould \& Goriely \cite{Arnould2006}). 
  The photodisintegration rates 
  $(\gamma,\alpha)$ are calculated using the reciprocity theorem (Eq. (4) of Arnould \& Goriely \cite{Arnould2006}).
  The resulting network is referred to as net0.

  \begin{table}[htbp]
	\caption{Initial conditions on the left- (column 2) and right-hand (column 3) sides of the discontinuity.}
	\label{table.ZND_godunov}
	\centering
	\begin{tabular}{lcc}
	\hline 
	\hline
	 & Left & Right \\
	 \hline  
	$\rho$ (g cm$^{-3}$) & $3.01 \times 10^8$ & $10^8$ \\
	$T$ (K) & $4.46 \times 10^{9}$ & $10^8$ \\
	$v$ (cm s$^{-1}$) & $8.07 \times 10^8$ & 0 \\
	\hline
	\hline
	\end{tabular}
  \end{table}
  
\begin{figure}
  \begin{center} 
	\resizebox{\hsize}{!}{\includegraphics{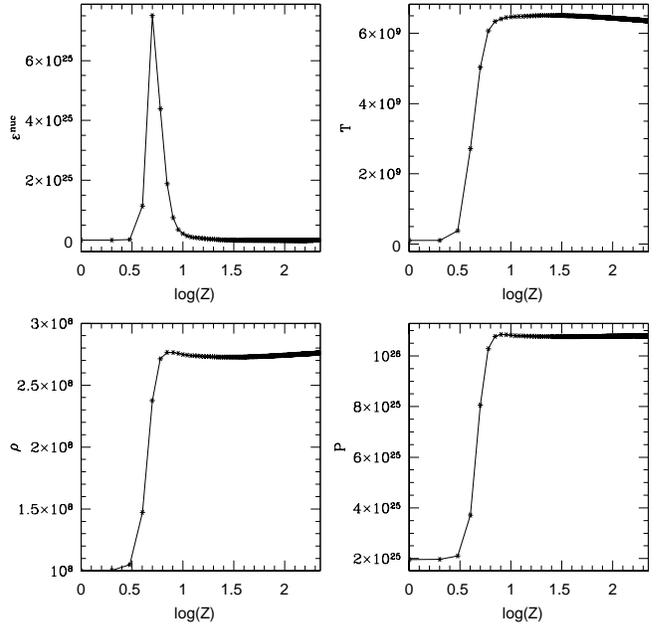}} 
  \caption{Nuclear energy generation (erg g$^{-1}$ s$^{-1}$), temperature (K), density (g$ \, $cm$^{-3}$) and 
  pressure (erg cm$^{-3}$) profiles for a detonation front in a mixture $X_{^{12}\mbox{C}}=0.2$, 
  $X_{^{96}\mbox{Ru}}=0.8$ at  $T=10^8 \, $K and $\rho= 10^8 \, $g$ \, $cm$^{-3}$. Z is the distance to the shock in cm. 
  }
  \label{therm2_C_80}
  \end{center} 
  \end{figure}

\begin{figure}
  \begin{center} 
        \resizebox{\hsize}{!}{\includegraphics{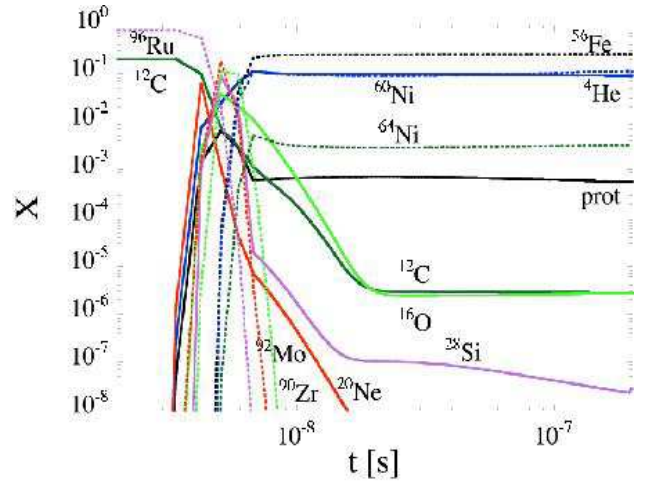}} 
  \caption{Nuclear mass fractions of some species calculated with the full network including 14758 
  reactions on 1381 nuclides using the profiles of Fig. \ref{therm2_C_80}. The time in s is the time elapsed 
  since the 
  passage of the shock, assuming a constant velocity $D = 1.19 \, 10^9$ cm s$^{-1}$.}
  \label{comp_steph}
  \end{center} 
  \end{figure}

\begin{figure}
  \begin{center} 
	\resizebox{\hsize}{!}{\includegraphics{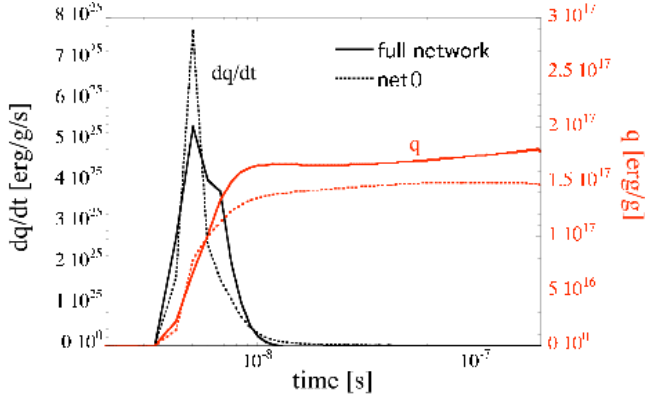}} 
  \caption{Network calculation of the energy production rate (erg g$^{-1}$ s$^{-1}$) and total energy 
  (erg g$^{-1}$) for a detonation in a mixture $X_{^{12}\mbox{C}}=0.2$, $X_{^{96}\mbox{Ru}}=0.8$ using the 
  profiles of fig. \ref{therm2_C_80} for two nuclear reaction network. Solid lines:  full 
  network including 14758 reactions on 1381 nuclides; dotted lines: reduced network net0. The time in s is measured since the 
  passage of the shock, assuming a constant velocity $D = 1.19 \, 10^9$ cm s$^{-1}$.}
  \label{therm_steph}
  \end{center} 
  \end{figure}

  \noindent  
  Using net0, we simulate a detonation, propagating in a mixture 
  $X_{^{12}\mbox{C}}=0.2$, $X_{^{96}\mbox{Ru}}=0.8$. The ignition conditions are 
  the same as used for pure $ ^{12}\mbox{C}$ in No\"el et al. (\cite{Noel2007}) (see Table \ref{table.ZND_godunov}). 
  The length of the domain is 1000 cm, the 
  initial discontinuity is placed at x = 100 cm, where x is the distance from the left boundary of the domain, and the 
  resolution is 1 cm (1000 numerical cells).
  We adopt a mixture $X_{ ^{56}\mbox{Ni}} = 0.1$ and $X_{ ^{64}\mbox{Ni}} = 0.9$, at the 
  left of the discontinuity. 
  The nuclear energy generation, temperature, density and pressure profiles, at time $t=7 \times 10^{-7}$ s 
  are presented in 
  Fig. \ref{therm2_C_80}, in terms of the distance to the shock.
  The detonation velocity is $D = 1.19 \times 10^9$ cm s$^{-1}$. 
  Using the thermodynamic profiles presented in Fig. \ref{therm2_C_80}, 
  we perform a full network calculation, including 14758 
  reactions on 1381 nuclides lying between the proton and neutron drip lines and with charge numbers Z$\leq$50. 
  Rates are taken from experiments, whenever available, and otherwise from the BRUSLIB library.
  The derived nuclear mass fractions, for some species are presented in Fig. \ref{comp_steph}.
  In Fig. \ref{therm_steph} we compare the energy production rates of the reduced network net0, and of the full network. 
  In this case the reduced network has reproduced the total energy production as well as the energy rate, 
  of the full network. This favorable situation may not however, be possible, for detonations propagating in
  a mixture composed of a lower amount of $^{12}$C.
 
\begin{figure}
  \begin{center} 
	\resizebox{\hsize}{!}{\includegraphics{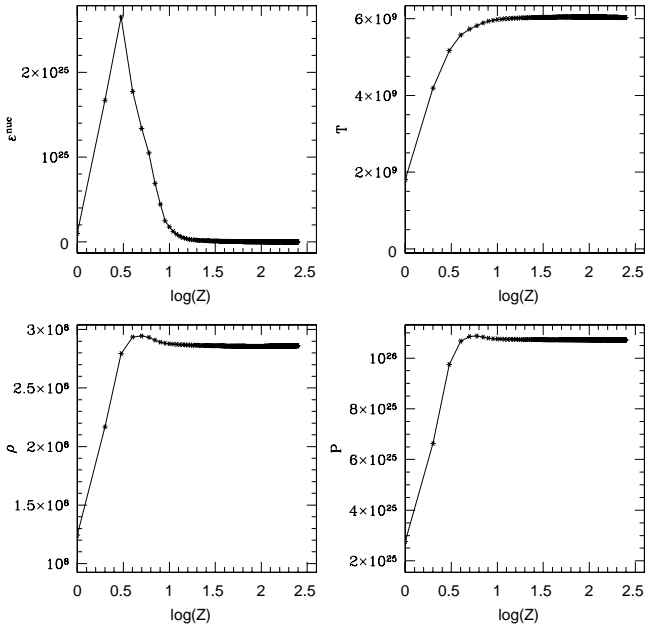}} 
  \caption{Same as Fig. \ref{therm2_C_80} but for a mixture  $X_{^{12}\mbox{C}}=0.1$, 
  $X_{^{96}\mbox{Ru}}=0.9$. 
  }
  \label{lgtherm_ru_09_r0}
  \end{center} 
  \end{figure}

\begin{figure}
  \begin{center} 
        \resizebox{\hsize}{!}{\includegraphics{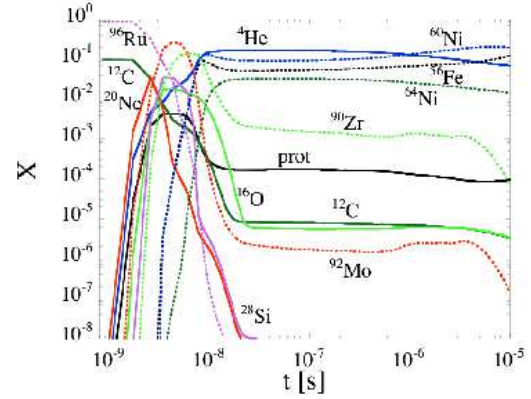}} 
  \caption{Same as Fig. \ref{comp_steph} but for a mixture  $X_{^{12}\mbox{C}}=0.1$, 
  $X_{^{96}\mbox{Ru}}=0.9$, using the profiles of Fig. \ref{lgtherm_ru_09_r0} and assuming a constant velocity 
  $D = 1.16 \, 10^9$ cm s$^{-1}$.}
  \label{claire_X_3D7}
  \end{center} 
  \end{figure}

  \noindent In a second step, we simulate a detonation propagating in a mixture 
  $X_{^{12}\mbox{C}}=0.1$, $X_{^{96}\mbox{Ru}}=0.9$. The ignition conditions are 
  similar to those used above.
  The nuclear energy generation, temperature, density, and pressure profiles, at time $t=7 \times 10^{-7}$ s 
  are presented in 
  Fig. \ref{lgtherm_ru_09_r0} in terms of the distance to the shock.
  The detonation velocity is $D = 1.16 \times 10^9$ cm s$^{-1}$. 
  Using the thermodynamic profiles of Fig. \ref{lgtherm_ru_09_r0}, a full-reaction network calculation is performed.
  The derived nuclear mass fractions, for some species, are presented in Fig. \ref{claire_X_3D7}.

\begin{figure}
  \begin{center} 
	\resizebox{\hsize}{!}{\includegraphics{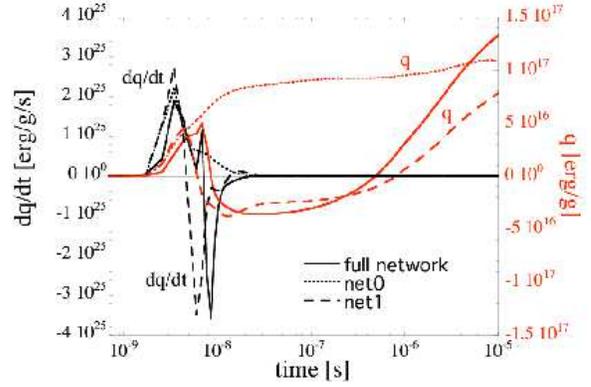}} 
  \caption{Same as Fig. \ref{therm_steph} but for a mixture $X_{^{12}\mbox{C}}=0.1$, $X_{^{96}\mbox{Ru}}=0.9$, 
  assuming a constant velocity $D = 1.16 \, 10^9$ cm s$^{-1}$. and for three nuclear reaction network. Solid lines:  full 
  network including 14758 reactions on 1381 nuclides; dotted lines: reduced network net0; dashed lines: 
  reduced network net1.}
  \label{claire_q_3D7}
  \end{center} 
  \end{figure}

  We compare the energy production rates, of the reduced network net0, and of the full network (Fig. \ref{claire_q_3D7}).
  We see that this case is characterized by an endothermic phase, due to the enhanced initial abundance of $^{96}$Ru, and 
  the significant contribution of the
  photodesintegration reactions, to the total energy balance, at an early stage. This phase of the energy generation 
  profile obtained using the full network, is not 
  reproduced by the reduced network net0.
  We therefore construct a more sophisticated reduced network, called net1. Let us describe the 
  methodology used in its construction. In the reduced networks, only the capture or emission of $\alpha$-particles are 
  taken into account. However, neutrons and protons can be produced significantly by photo-reactions, and be
  recaptured. In particular, the nucleus (Z-2,N-2), where Z is the charge number, and N the number of neutrons, 
  can be produced by the photodesintegration of the nucleus (Z,N), following 
  six additional paths, complementing the $(\gamma, \alpha)$ one; these include the emission of two neutrons and 
  two protons.
  The characteristic timescale, for each of the seven possible paths is given by

  \begin{eqnarray}\label{eq.lambda}
  \lambda_0^{-1} & = & \lambda_{(\gamma,\alpha)}^{-1} \mbox{,} \nonumber \\
  \lambda_1^{-1} & = & \lambda_{(\gamma,n)}^{-1}(Z,N)+\lambda_{(\gamma,n)}^{-1}(Z,N-1)+\lambda_{(\gamma,p)}^{-1}(Z,N-2) \nonumber \\
                 &   & +\lambda_{(\gamma,p)}^{-1}(Z-1,N-2) \mbox{,} \nonumber \\
  \lambda_2^{-1} & = & \lambda_{(\gamma,n)}^{-1}(Z,N)+\lambda_{(\gamma,p)}^{-1}(Z,N-1)+\lambda_{(\gamma,n)}^{-1}(Z-1,N-1) \nonumber \\
                 &   & +\lambda_{(\gamma,p)}^{-1}(Z-1,N-2) \mbox{,} \nonumber \\
  \lambda_3^{-1} & = & \lambda_{(\gamma,n)}^{-1}(Z,N)+\lambda_{(\gamma,p)}^{-1}(Z,N-1)+\lambda_{(\gamma,p)}^{-1}(Z-1,N-1) \nonumber \\
                 &   & +\lambda_{(\gamma,n)}^{-1}(Z-2,N-1) \mbox{,} \nonumber \\
  \lambda_4^{-1} & = & \lambda_{(\gamma,p)}^{-1}(Z,N)+\lambda_{(\gamma,n)}^{-1}(Z-1,N)+\lambda_{(\gamma,n)}^{-1}(Z-1,N-1) \nonumber \\
                 &   & +\lambda_{(\gamma,p)}^{-1}(Z-1,N-2) \mbox{,} \nonumber \\
  \lambda_5^{-1} & = & \lambda_{(\gamma,p)}^{-1}(Z,N)+\lambda_{(\gamma,n)}^{-1}(Z-1,N)+\lambda_{(\gamma,p)}^{-1}(Z-1,N-1) \nonumber \\
                 &   & +\lambda_{(\gamma,n)}^{-1}(Z-2,N-1) \mbox{,} \nonumber \\
  \lambda_6^{-1} & = & \lambda_{(\gamma,p)}^{-1}(Z,N)+\lambda_{(\gamma,p)}^{-1}(Z,N-1)+\lambda_{(\gamma,n)}^{-1}(Z-2,N) \nonumber \\
                 &   & +\lambda_{(\gamma,n)}^{-1}(Z-2,N-1) \mbox{,}
  \end{eqnarray}
  so that the total photo-reaction rate of (Z,N), leading to (Z-2,N-2), can be approximated by the effective rate
  \begin{equation}\label{superpo}
  \lambda_{eff}=\lambda_0+\lambda_1+\lambda_2+\lambda_3+\lambda_4+\lambda_5+\lambda_6 \mbox{.}
  \end{equation}
  The reverse $\alpha$-capture rate can then be calculated using the detailed balance equation (Eq. (4)
  of Arnould \& Goriely \cite{Arnould2006}).
  While $\lambda_0$, describing $\alpha$-photoemission, is the only path included in the net0 approximation, 
  the neutron and/or proton photoemission could potentially dominate in the case of light (Z$<$40) species, depending on the nuclei 
  involved. In the neutron-deficient region (especially along the neutron-shell closures), the proton-photoemission is dominant, while close to the stability line 
  the neutron-photoemission usually dominates. Any of these seven paths, however,
  can be impeded to a more or less large extent
  by the reverse nucleon captures that would modify 
  the relative significance of a given path.
  This modification
  depends on the thermodynamic and composition conditions, as well 
  as on the nuclear cross sections. To 
  determine which paths are important, we use the following method. We first calculate the detonation profiles 
  using net0. On the basis of the temperature and density profiles (for instance those of Fig. \ref{lgtherm_ru_09_r0} for 
  $X_{^{12}\mbox{C}}=0.1$, $X_{^{96}\mbox{Ru}}=0.9$), the 
  full reaction network 
  is solved. This calculation allows to determine the dominant 
  paths and to construct  
  a reduced network with new effective reaction rates but the same reduced number of nuclei which is introduced in the hydrodynamical algorithm. We perform the detonation simulation 
  using this new reduced network, and the resulting thermodynamical profiles are used again in a full network calculation. 
  We iterate between a hydrodynamical simulation with a given reduced network, and the readjustment of 
  a new reduced network, on the basis of the previously-derived hydrodynamical profiles, until the reduced 
  and full 
  networks converge to achieve high accuracy in the energy production rate.
  Note that no strict mathematical criteria have been used to estimate the quality of the newly-developed 
  reaction network. 
  In contrast, we  estimate that the new network must fulfill two major conditions: it must reproduce the 
  right 
  order of magnitude (roughly within a factor of two) of the total amount of energy produced, and the  possible 
  appearance of  endothermic phases. 
  Indeed any endothermic phase can play a role of first importance on the detonation 
  properties. In particular pathological detonations may occur
  (Fickett \& Davis \cite{Fickett1979}, Khoklov \cite{Khoklov1989}, Sharpe \cite{Sharpe1999}).

  In contrast to the reduced network used in No\"el et al. (\cite{Noel2007}), which provided an 
  energy source typical of explosive helium and carbon burning in the absence of hydrogen 
  (Gamezo et al. \cite{Gamezo1999}), our reduced network  
  depends on the detonation profiles, and initial composition. For each new detonation simulation, we construct the 
  adapted reduced network using the iteration procedure described above.
  
  Convergence is achieved in the case of $X_{^{12}\mbox{C}}=0.1$, $X_{^{96}\mbox{Ru}}=0.9$, with the following 
  reduced network, called net1, in which the equivalent reaction rate, adopted for each photodisintegration from $^{92}$Mo 
  to $^{64}$Ni, and 
  from $^{56}$Ni to $^{16}$O is 
  \begin{equation}\label{eq.lambanet1}
  \lambda_{net1}=\lambda_0+\lambda_1+\lambda_6 \mbox{,}
  \end{equation}
  and for $^{96}$Ru, $\lambda_{net1}=\lambda_0$.
  The inverse $\alpha$-capture rates are calculated using Eq. (4) of Arnould \& Goriely (\cite{Arnould2006}), and the five remaining rates for 
  $ ^{12}$C($ ^{12}$C,$\alpha$)$ ^{20}$Ne, $ ^{12}$C($ ^{16}$O,$\alpha$)$ ^{24}$Mg, $ ^{16}$O($ ^{16}$O,$\alpha$)$ ^{28}$Si, 
  $ ^4$He$(2\alpha,\gamma)$$ ^{12}$C, and $ ^{12}$C$(\gamma,2\alpha)$$ ^4$He, are similar to those used in net0.
  
\begin{figure}
  \begin{center} 
	\resizebox{\hsize}{!}{\includegraphics{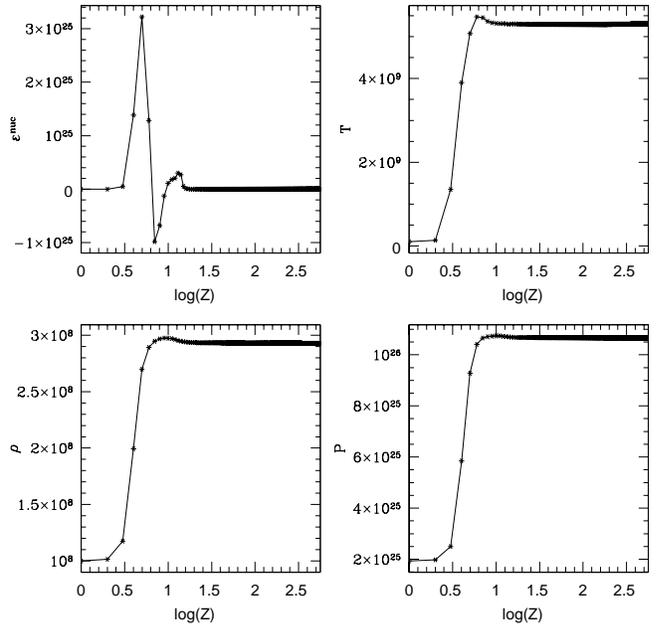}} 
  \caption{Same as Fig. \ref{lgtherm_ru_09_r0}, but with net1.
            }
  \label{lgtherm_r0_r1_r6}
  \end{center} 
  \end{figure}

  Figure \ref{claire_q_3D7} compares the energy production calculated using net0, net1, and the full network. Using the profiles obtained with net0 
  (Fig. \ref{lgtherm_ru_09_r0}), we perform a full network calculation, and compute the energy production rate. 
  The same profiles are used for the reduced network net0, and the energy production rates are compared. We see 
  that the endothermic part of the profile is not reproduced by the reduced network net0. A revised reduced network is 
  therefore developed,
  net1, which simulates more accurately, the full network energy production rate.
  However, new detonation
  profiles, using the hydrodynamical algorithm have to be calculated with this new network.
  They are shown in Fig. \ref{lgtherm_r0_r1_r6}, at time $t_f=1.6 \times 10^{-6}$s. The domain length 
  is of 2000 cm, and the resolution is of 1 cm. The initial discontinuity is positioned at x = 200 cm. The detonation
  velocity is $D=1.09 \times 10^9$ cm s$^{-1}$.
  
\begin{figure}
  \begin{center} 
	\resizebox{\hsize}{!}{\includegraphics{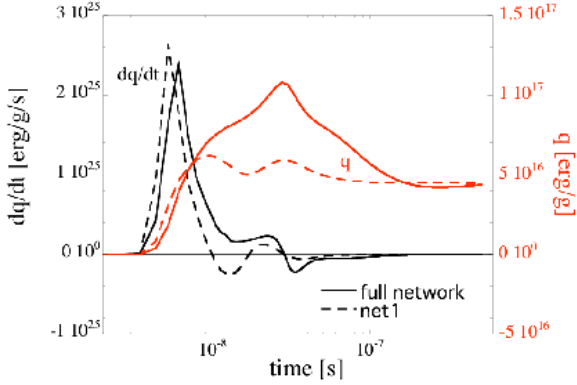}} 
  \caption{Same as Fig. \ref{claire_q_3D7}, but with the profiles of Fig. \ref{lgtherm_r0_r1_r6} and  
  $D=1.09 \, 10^9$ cm s$^{-1}$.}
  \label{claire_q_3D9}
  \end{center} 
  \end{figure}

\begin{figure}
  \begin{center} 
	\resizebox{\hsize}{!}{\includegraphics{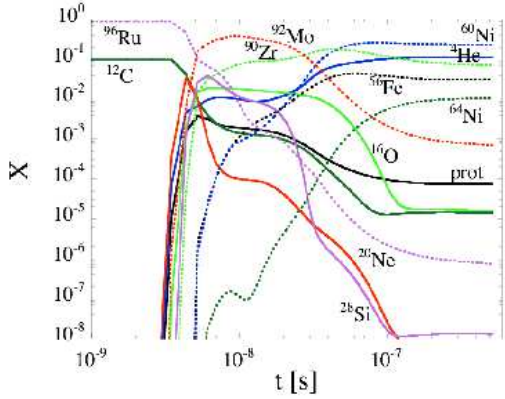}} 
  \caption{Same as Fig. \ref{claire_X_3D7} but with the profiles of Fig. \ref{lgtherm_r0_r1_r6} and  
  $D=1.09 \, 10^9$ cm s$^{-1}$.
  }
  \label{claire_X_3D9}
  \end{center} 
  \end{figure}

  Using these new thermodynamic profiles (Fig. \ref{lgtherm_r0_r1_r6}), network calculations of the energy production are made for 
  the full network, and the reduced network net1, which are compared in Fig. \ref{claire_q_3D9}. One sees that 
  the reduced network reproduces satisfactorily the energy production given by the full network. We can therefore 
  consider net1 as
  a suitable reduced network for the simulation of a detonation for a mixture $X_{^{12}\mbox{C}}=0.1$, 
  $X_{^{96}\mbox{Ru}}=0.9$, and the initial conditions shown in Table \ref{table.ZND_godunov}. 
  Adopting the thermodynamic profiles shown in Fig. \ref{lgtherm_r0_r1_r6}, 
  a final calculation is performed using the full network. 
  The derived nuclear mass fractions for some species, are presented in Fig. \ref{claire_X_3D9}.
  A comparaison between Fig. \ref{claire_X_3D7} and Fig. \ref{claire_X_3D9}, reveals the sensitivity of the 
  final composition to the choice of nuclear reaction network. For instance, net0 produces much less
  $^{90}$Zr and $^{92}$Mo than net1, at the end of the reaction zone.
  
\begin{figure}
  \begin{center}
  	\resizebox{\hsize}{!}{\includegraphics{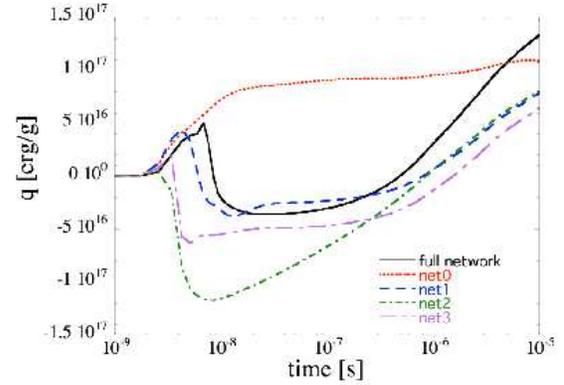}} 
  \caption{Same as Fig. \ref{claire_q_3D7} but for five different networks and only the total energy produced 
  is presented. The network net2 is identical to net0 for all species between $^4$He and $^{56}$Ni, 
  and uses $\lambda_{eff}$ for the rates of all reactions between $^{64}$Ni and $^{96}$Ru. The network net3 
  uses $\lambda_{eff}$ for the rates of all photo-reactions between $^{16}$O and $^{96}$Ru, 
  and the net0 rates for the other reactions.
  }
  \label{claire_q_3D7b}
  \end{center} 
  \end{figure}

  Far more than two reduced networks were tested in the iteration procedure. Figure
  \ref{claire_q_3D7b} displays
  the total energy obtained, using five different reduced networks, for the thermodynamic profiles 
  of Fig. \ref{lgtherm_ru_09_r0}. It is clear that net1 reproduces most closely the global energy
  obtained by the full network simulation.

  Note that net1 includes photoneutron and photoproton paths that are found to reproduce globally 
  all possible paths, when including neutrons and protons in the network. Both paths (path 1 and path 6 in 
  Eq. \ref{eq.lambda}) are required to simulate the bottlenecks which are related to the neutron (N=20, 28, 50), or 
  proton (Z=20, 28) magic 
  numbers, which are crossed by the photonuclear flow. 
  All nuclei included in the reduced network are even-even, such that if a first $(\gamma,n)$ reaction is possible, 
  the resultant (Z,N-1) nucleus will certainly photo-emit a neutron. The same holds for 
  proton emission. The equivalent rate $\lambda_{net1}$ in Eq. \ref{eq.lambanet1}, simulates the 
  possible inclusion of neutrons and protons in the network. The inverse effective $\alpha$-captures rate, 
  that we deduced from detailed balance expressions,
  similarly enable to simulate the possible neutron and proton captures, without having 
  to include neutrons and protons explicitely in the network.
  Finally note that for $^{96}$Ru, we do not take the ($\gamma$,n) and ($\gamma$,p) contributions from 
  Eq. \ref{eq.lambanet1}, into account to counterbalance the neutron and proton captures by $^{96}$Ru, at
  early time when the matter is almost entirely made of $^{96}$Ru.
 
\begin{figure}
  \begin{center} 
	\resizebox{\hsize}{!}{\includegraphics{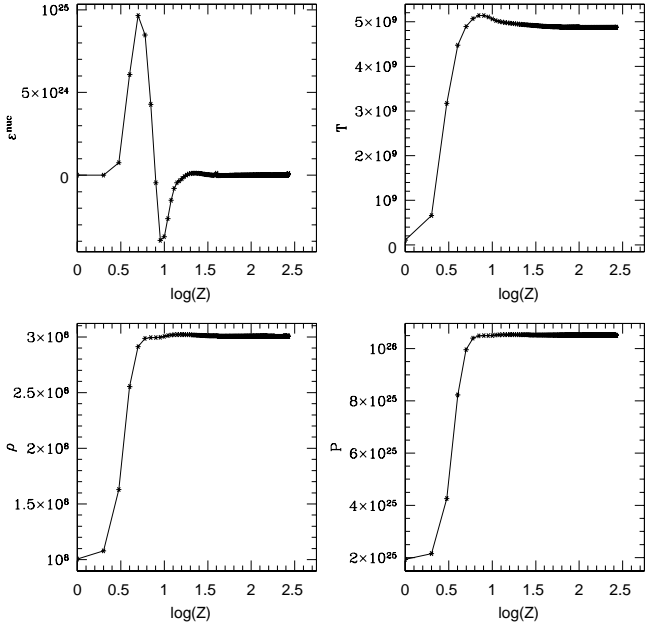}} 
  \caption{Same as Fig. \ref{lgtherm_ru_09_r0} but with $X_{^{12}\mbox{C}}=0.05$ and
	$X_{^{96}\mbox{Ru}}=0.95$.}
  \label{lgtherm_r0_r1_r6_95pc}
  \end{center} 
  \end{figure}

\begin{figure}
  \begin{center} 
	\resizebox{\hsize}{!}{\includegraphics{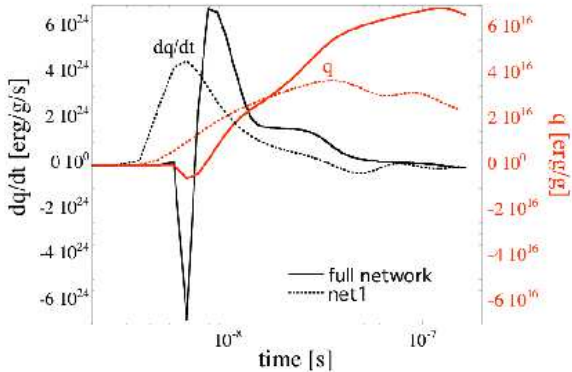}} 
  \caption{Same as Fig. \ref{claire_q_3D7} but with the profiles of Fig. \ref{lgtherm_r0_r1_r6_95pc},  
  $D=9.88 \, 10^8$ cm s$^{-1}$ and $X_{^{12}\mbox{C}}=0.05$, $X_{^{96}\mbox{Ru}}=0.95$.}
  \label{claire_q_3D2_95pc}
  \end{center} 
  \end{figure}

  We emphasize that the reduced network considered for 
  the detonation of the  mixture of $X_{^{12}\mbox{C}}=0.1$, $X_{^{96}\mbox{Ru}}=0.9$ cannot be generalized  automatically 
  to cases characterized by different initial conditions.  The construction of the reduced network does not only depend 
  on the  temperature and density time evolution, but also on the specific  initial composition. Different initial mass 
  fractions of the existing species, or a different elemental or isotopic composition, would require the construction 
  of a new network, with new optimum effective reaction rates. 
  As an example, we present a second case study, in which $X_{^{12}\mbox{C}}=0.05$, 
  $X_{^{96}\mbox{Ru}}=0.95$, as considered by Cumming \& Bildsten (\cite{cumming2001}).
  The detonation profiles, in this case, are shown in Fig. \ref{lgtherm_r0_r1_r6_95pc}, using net1. The domain length 
  is 1000 cm, the resolution is 1 cm, the initial discontinuity is placed at x = 100 cm, and the detonation
  velocity is $D=9.88 \times 10^8$ cm s$^{-1}$.
  We see in Fig. \ref{claire_q_3D2_95pc} that net1 does not reproduce the endothermic phase. 
  A more suited network needs to be defined, for example, by considering a linear superposition of photorates, 
  as in Eq. \ref{superpo}, that is different for each nucleus included in the network. Wheter such a scheme
  should be developed depends on the outcome of realistic modelling, of the initial composition of superbursts,
  in heavy elements. 

\section{Conclusion}\label{Conclusions}

  We have shown that a simple extension of an $\alpha$-chain, might not be sufficiently accurate to reproduce the
  energy production rate of a detonation in a mixture that contains too much heavy elements.
  We have developed a new methodology to construct a reduced network, with adapted effective reaction rates, based on
  an iteration procedure between hydrodynamical simulations, with a chosen reduced network, and a comparison with full 
  network calculations.
  In particular, we have constructed a new reduced network that reproduces satisfactorily, the energy generation rate of a 
  reference network,
  including all reactions on all nuclides, probably involved in the specific case of the propagation of a detonation 
  at $T=10^8$K, and $\rho=10^8$ g cm$^{-3}$, in a mixture 
  $X_{^{12}\mbox{C}}=0.1$, $X_{^{96}\mbox{Ru}}=0.9$.
  Adopting such a reduced network, we have simulated detonation profiles.
  These profiles are used in a full network 
  simulation to compute the associated nucleosynthesis. One sees that the presence of $^{96}$Ru decreases the total 
  energy of the detonation, due to an endothermic phase at early stages. For this reason, the propagation velocity of the detonation decreases.
  For $X_{^{12}\mbox{C}}=0.1$, 
  $X_{^{96}\mbox{Ru}}=0.9$,  
  essentially only iron-group elements remain at the end of the reaction zone, with some addition of N=50 elements, such as 
  $^{90}$Zr. This fact can 
  influence the neutron star crust composition and properties, such as its conductivity, or its neutrino emissivity, 
  as well as the ignition conditions for the superbursts 
  (Weinberg \& Bildsten \cite{Weinberg2007}, Gupta et al. \cite{Gupta2007}, Haensel \& Zdunik \cite{Haensel1990}).
  
  Interestingly, it can be seen that for $X_{^{96}\mbox{Ru}}>0.9$, the detonation goes through a short endothermic 
  phase (see Figs. 
  \ref{lgtherm_r0_r1_r6}-\ref{claire_q_3D2_95pc}) that cannot be described by net0 
  (Fig. \ref{lgtherm_ru_09_r0}). 
  The photodesintegrations by $\alpha$ emission are endothermic reactions, and
  only the recombination of $\alpha$ particles up to Ni compensates for the energy loss. 
  The energy is therefore released by the captures or recombinations of the protons, neutrons and 
  $\alpha$-particles, liberated by the photodesintegrations. 
  Furthermore a comparaison between Fig. \ref{lgtherm_ru_09_r0} and Fig. \ref{lgtherm_r0_r1_r6} reveals that the 
  temperature reached during the 
  detonation is lower using net1, than net0, which influences the nucleosynthesis.
  
  Computationally mixed C/Ru detonations are relatively easy to handle since the length-scales on which 
  the different elements burn are comparable. This contrasts with the case of pure carbon detonations 
  where a large variety of length-scales is obtained (No\"el et al. \cite{Noel2007}), and where multiple simulations at different resolutions 
  are required to fully resolve the detonation. In the C/Ru case, carbon and ruthenium burn on 
  $\approx$ 10 cm, and the total reaction length is $\approx$ 1000 cm. The length-scales span only three 
  orders of magnitude compared to six for pure carbon detonations. This is an encouraging feature for future 
  multi-dimensional calculations.

\begin{acknowledgements}

      The numerical simulations presented herein were performed on the parallel computers of the Intensive Computing Storage 
      of UCL and on HYDRA, the Scientific Computer Configuration at the VUB/ULB Computing Centre. 
      We are grateful to M. Arnould for interesting discussions.

\end{acknowledgements}

\end{document}